\documentstyle[jkas]{article}

\beginpage{1}
\endpage{4}
\year{2010}\volume{43}\month{February}

\runningauthor {F.Shi et al.}
\runningtitle{Star Formation Rate Indicators in WISE/SDSS}

\month{February} \year{2010} \volume{43} \issueno{4}
\beginpage{125}\endpage{130}
\date{Received December 1, 2009; Accepted January 17, 2010}

\begin{document}
\newcommand{\rmn}[1] {{\rm #1}}
\newcommand{\Zsolar}{\mbox{${\; {\rm {\sun}}}$}}
\newcommand{\ha}{\hbox{H$\alpha$}}
\newcommand{\oii}{\hbox{[O\,{\sc ii}]}}
\newcommand{\neiii}{\hbox{[Ne\,{\sc iii}]}}
\newcommand{\siii}{\hbox{[S\,{\sc iii}]}}
\newcommand{\ariii}{\hbox{[Ar\,{\sc iii}]}}
\newcommand{\nii}{\hbox{[N\,{\sc ii}]}}
\newcommand{\sii}{\hbox{[S\,{\sc ii}]}}
\newcommand{\oiii}{\hbox{[O\,{\sc iii}]}}
\newcommand{\hb}{\hbox{H$\beta$}}

\title{Star Formation Rate Indicators in WISE/SDSS}
\author{Fei Shi$^{1}$, Xu Kong$^{2,3}$, Yu-Yan  Liu$^{1}$, Shan-Shan  Su$^{2}$,
Yang Chen$^{2}$, Jin-Rong Li$^{2}$, Zi-Qiang Gong$^{1}$, and Dong-Xin Fan$^{4}$   }
\address{$^1$ North China Institute of Aerospace Engineering, Langfang,
Hebei, 065000, China.\\
 {\it E-mail : fshi@bao.ac.cn}}
\address{$^2$ Center of Astrophysics, University of Science and
Technology of China, Jinzhai Road 96, Hefei 230026, China.\\
{\it E-mail : xkong@ustc.edu.cn}}
\address{$^3$ Key Laboratory for Research in Galaxies and Cosmology,
University of Science and Technology of China, CAS, China.\\
{\it E-mail : xkong@ustc.edu.cn}}
\address{$^4$ Department of Physics and Electronics, Guangxi Teachers
Education University, Nanning 530001, China.\\
{\it E-mail : fandx3280@126.com}}
\address{\normalsize{\it (Received July 1, 2013; Accepted July **, 2013)}}
\offprints{F.Shi}
\abstract{With the goal of investigating the degree to which the total infrared
luminosity ($L_{\rm TIR}$) traces the star formation rate (SFR),
we analyze the $L_{\rm TIR}$ from the dust   in a sample of $\sim$
6000 star-forming galaxies, based on
the 3.4, 4.6, 12 and  22 $\mu$m data from the Wide-field Infrared Survey Explorer (WISE)
 and u, g, r, i, z band data from SDSS DR9. These star-forming galaxies  are selected
by matching the WISE All Sky Catalog with the star-forming
galaxy catalog in SDSS DR9 provided by
JHU/MPA\thanks{$http://www.sdss3.org/dr9/spectro/spectroaccess.php$}.
The values of $L_{\rm TIR}$ and SFR are derived from the project Code Investigating Galaxy Emission
(CIGALE). We study the relationship between the $L_{\rm TIR}$ and SFR. From this study, we
derive reference SFR indicators for use in our analysis.
Linear correlations between SFR and the $L_{\rm TIR}$
are found, and calibrations of SFRs based on it are proposed.
The calibration holds for galaxies with verified  observations and
agrees well with previous works. The dispersion in the
relation between $L_{\rm TIR}$  and SFR could partly be explained by
the galaxy's properties, such as the 4000 {\AA} break.}

\maketitle

\section{INTRODUCTION}
The star formation rate (SFR) is a crucial parameter to characterize
the star formation history of galaxies.
To calculate the SFR reliably, extensive efforts have been made to
derive SFR indicators at various wavelengths, including radio, infrared
 (IR), ultraviolet (UV), optical spectral lines and continuum
(Kennicutt 1998).

 UV SFR indicators can directly probes the bulk of the emission from young,
massive stars, which make it as a good SFR tracer, while it is highly sensitive to dust
reddening and attenuation whose correction techniques depends on the nature of the galaxy
( Buat et al. 2002, 2005, Salim et al. 2007).
The optical and NIR SFR indicators, such as
hydrogen recombination lines (\ha,\hb) and
forbidden line emission (\oii, \oiii) are also good tracers of the ionizing photons
from young, massive stars, while it is affected by not only  dust extinction
though to a much lesser degree than the UV, but also impacted by assumptions on
the underlying stellar absorption and on the form of the high end of the stellar initial mass
function(Kong et al. 2004; Calzetti et al. 2005).
Infrared SFR indicators can be complementary to UV and optical indicators
because they measure star formation via the dust-absorbed stellar light
that emerges beyond a few $\mu$m(da Cunha et al. 2008).
As a result, SFR indicators in the infrared (IR) band have
attracted more attention in recent years because of the data with high sensitivity
and high angular resolution provided by the Spitzer Space Telescope.
As a result, the general correlation between infrared luminosity and SFR
has been found and calibrated (Calzetti et al.  2007, 2009, 2010,
AlonsoHerrero et al. 2006, Moustakasal et al. 2006, Schmitt et al. 2006,
Kennicutt et al. 2007, Persic et al. 2007, Rosa et al. 2007, Bigiel et al. 2008, Rieke et al. 2009).

The  monochromatic (i.e., one band or wavelength) infrared SFR diagnostics using
Spitzer IRAC detections at 8$\mu$m, Spitzer MIPS detections at 24$\mu$m, 70$\mu$m, 160$\mu$m removes the need
for highly uncertain extrapolations of the dust spectral energy distribution across the full
wavelength range while it suffers losing most of information in the UV-optical-IR wavelength range
(Calzetti et al.2010, Rujopakarn et al.2013).

During the last twenty years, many calibrations of these correlations have
 focused on the relation between the total luminosity in the IR
band ($L_{\rm TIR}$) and SFR because of dust heating in  the wide IR
band. Calculation of $L_{\rm TIR}$ requires models describing the infrared spectral
energy distribution (SED) of star-forming galaxies (Chary \& Elbaz 2001,
 Dale \& Helou 2002, Lagache et al.2003, Marcillac et al.2006),
 but these calibrations usually suffer from small sample size for galaxies
and limited sensitivities from surveys such as the Infrared Astronomical
Satellite (IRAS) and Infrared Space Observatory (ISO).

The Sloan Digital Sky Survey (SDSS) is the most ambitious
imaging and spectroscopic survey to date, and have
covered 14,555 square degrees  of the sky in the ninth SDSS Data Release.
The large area coverage and moderately deep survey limit of the SDSS make it
suitable to study the large-scale structure and the characteristics
of the galaxy evolution(York et al. 2000).

The Wide-field Infrared Survey Explorer (WISE, Wright et al. 2010) has
 mapped the entire sky with 5 $\sigma$ point source sensitivities better
than 0.08, 0.11, 1 and 6 mJy in bands centered on wavelengths of 3.4, 4.6, 12 and 22
$\mu$m respectively, which is much more sensitive than
previous surveys (Cutri et al. 2012).
For example, WISE is achieving 100 times better sensitivity than IRAS
in the 12 $\mu$m band. As an all-sky survey, WISE has returned data for
 over 9.4 billion objects, so together with SDSS, WISE provides us with a large sample
of star-forming galaxies  having  reliable photometric flux measurements.
The improved sensitivity and large sample size make it suitable for
studying the evolution of galaxies.

Start from building a star-forming sample with multi-wavelength observation flux,
we derive the SFR for each
galaxy by the spectral energy distribution (SED) fitting. $L_{\rm TIR}$ were correlated with the
SFR to built the SFR calibrations. This paper is organized as follows. Based on the WISE All Sky  Data
Release, we present a sample to derive our SFR index calibration
(Sect. 2). In Sect. 3, we study the correlation between SFR and the
 $L_{\rm TIR}$  and perform the numerical calibration.
 In Sect. 4, we summarize the calibration result and conclude this
paper. Throughout the paper, we adopt cosmological parameters of the $\Omega_M$ = 0.27
and $\Omega_\Lambda$ = 0.73.

\section{Data sample selection}

  To calibrate the relation between $L_{\rm TIR}$  and SFR,
   we have to first generate a sample from  a catalog of sources that have good detections from the WISE All Sky Survey.
  All the objects in the sample must have reliable flux measurement and therefore
  reliable calculation of SFR and $L_{\rm TIR}$.
  For this purpose, we set the selection criteria as follows.
  \begin{enumerate}
\item
The source must have a valid detection and the detection must have a signal to noise ratio
$(SNR)>5$ for 3.4, 4.6, 12 and  22 $\mu$m bands.
\item
 In order to eliminate spurious detections in low coverage areas within Atlas Tiles,
 such as cosmic ray strikes, residual satellite trails, hot pixel events and scattered light
 from very bright moving objects such as the moon and planets, we require
 a reliable band detection which must have been extracted from a region with
 at least five independent frames image
and at least two of the frame image must have been detected with $SNR>3$.
\item
 {\bf In order to eliminate entries that may have incorrect aperture photometry in the largest measurement aperture,
 which has a radius of 50",  we require the distance to an Atlas Image boundary
 in the range of  between 36.4 and 4058.6 because Atlas Images are 4095x4095 pixels in size and the coadd pixel scale is 1.375"/pix.}
\item
Filter out spurious detections of image artifacts caused by bright sources.
 \item
  We are only interested in extragalactic  galaxies, so we exclude galactic latitudes between -10.0 degrees and 10.0 degrees
  which are dense with stars in the Milky Way.
\item
{\bf A variable source is not allowed in our selected galaxy sample by requiring the variability flag($var_{flg}$)
that is related to the probability of  that the source is variable in that band,  to be values of "0" through "4" .}
\item
Ultra-luminous galaxies beyond $z\sim0.1$ have very different properties compared to the low-redshift star-forming galaxies
   and hence removed from our sample(Rujopakarn et al.2013).
\end{enumerate}

   After compiling a catalog of sources that have good detections from the WISE All-sky Survey,
 we match this  catalog  with the catalog of star-forming galaxies in SDSS
   DR 9 provided by MPA which made use of
the spectral diagnostic diagrams from Kauffmann et al. (2003)
to classify galaxies as starburst galaxies, active galactic nuclei
(AGN), or unclassified. A star-forming galaxy is selected by requiring arc distances
   between the galaxy identified in SDSS and WISE Catalog
   to be less than the 6.1" which is the angular resolution of 3.4 $\mu$m bands.

    {\bf In total, 5,930 star-forming galaxies are adopted in our sample and
 no multiple matches are found. We use the photometric redshifts  provided by SDSS DR9.
 The flux of WISE in 3.4, 4.6, 12 and  22 $\mu$m bands is  aperture curve-of-growth corrected
 and measured in an   circular aperture centered on the source position on the Atlas Image.
  The flux of SDSS in u, g, r, i, z band is aperture corrected by taking the best fit exponential
  and de Vaucouleurs fits in each band and asks for the linear combination of the two that best fits the image.}
  The estimation of k-corrections for the photometry  observations is not necessary because
 the redshifts of  the galaxies in the sample is restricted to be less than 0.1.

\section{ Calculation of Star Formation Rate and $L_{\rm TIR}$ }

An efficient way to derive physical parameters of star formation
 homogeneously is to fit the observed
SED with models from a stellar population synthesis code.
We use the  SED fitting program package CIGALE (Code Investigating Galaxy
Emission, Noll et al. 2009 and  Serra et al. 2011) to derive $L_{\rm TIR}$  and SFR.
CIGALE was developed to derive highly reliable galaxy properties by fitting the
 UV-to-far-IR SED and the related dust emission at the same time, i.e., the stellar population
synthesized models are connected with infrared templates by the balance of the energy of dust
emission and absorption. We refer the reader to Noll et al. (2009) for a detailed description.

{\bf For CIGALE, all the light from stars is assumed to be absorbed by dust,
  and re-emitted in the IR(Buat et al. 2011).
  SFR is the instantaneous SFR, defined as $(1-f_{ySP})*SFR_1+f_{ySP}*SFR_2$,
where $f_{ySP}$ is the burst strengths(mass fraction of young stellar population), $SFR_1$ and $SFR_2$ correspond
to the star formation rates of the old and young stellar populations, respectively.}

Firstly, we {\bf combine} the redshift-dependent filter fluxes of
the 3.4, 4.6, 12 and  22 $\mu$m data from WISE and model flux from SDSS DR9 for our sample, which
having been aperture corrected.

Secondly, we apply different scenarios for young and old populations in the same way as in Buat et al. (2011).
The input star formation history (SFH) here is a constant burst SFH for young stellar populations,
and an exponentially decreasing one for old stellar populations.
The stellar population synthesis models of Maraston (2005) are adopted in CIGALE,
which include a full treatment of the thermally pulsating asymptotic giant branch (TP-AGB) stars.
A Salpeter IMF is used to calculate the complex stellar populations and
the stellar metallicity is taken to be solar.

Thirdly, we choose the attenuation curve, which is  based on a law given by Calzetti et al. (2000),
with modification of the slope and/or by adding a UV bump.
The modification of the slope is controlled by the factor $(\lambda/\lambda_{V})^{\delta}$,
where $\lambda_{V}$ = 5500 ${\AA}$ is the reference wavelength of the V filter,
and  $\delta$ is the slope of the attenuation curve.
We only considered a modification of the slope for the attenuation law here, and no UV bump was introduced
 because {\bf  there is no clear evidence of the presence of such a UV bump for the local galaxies
 (Buat et al. 2011, and reference therein).
  We consider different effects of attenuation for old and young stellar populations
  by adding the reduction factor $f_{\rm att}$ of V-band attenuation for old stellar population(relative to the young one),
   describing dust attenuation for the old stellar populations.

Fourthly,  we choose IR power-law slope $\alpha$, which relates the dust mass to heating intensity.
The  parameter $\alpha$ is in the interval [1 ;2.5] to cover the main domain of activity.}

Finally, depending on the $\chi^2$ for the best-fitting
models with a set of bins in the parameter space, probability distributions
as a function of the parameter value are calculated and
used to derive weighted galaxy properties, such as $L_{\rm TIR}$  and instantaneous SFR.
The input parameters listed in Table 1 have been demonstrated to be able to obtain a stable and reliable output.

\begin{table*}
 \centering
\begin{tabular}{ l c l}

  \hline
  {Parameters} & {Symbol} & {Range} \\
    \hline
 {Star formation history} & &   \\
  \hline
metallicities (solar metallicity) & $Z$ &  0.02 \\
 $\tau$ of old stellar population models in Gyr & $\tau_1$ & 0.1;  1.0; 10.0 \\
ages of old stellar population models in Gyr & $t_1$ & 2.0;5  \\
ages of young stellar population models in Gyr & $t_2$ &  0.05;0.2\\

fraction of young stellar population & $f_{ySP}$ & 0.001; 0.01; 0.1; 0.999 \\
IMF & S & Salpeter \\

  \hline
  {Dust attenuation} & &  \\
  \hline
  Slope correction of the Calzetti law & $\delta$ &  -0.2; -0.1; 0.0; 0.1; 0.2\\
  V-band attenuation for the young stellar population & $A_{V,ySP}$ &  0.15; 0.3; 0.45; 0.60; 0.75 ;0.90 ;1.05 ;\\
                                                                   & &  1.20;1.35;1.50;1.65;1.80;1.95;2.1  \\

  Reduction of $A_{V}$  based on the old SP model & $f_{\rm att}$ & 0.50;1.0  \\
    \hline
 {IR SED} & &\\
    \hline
 IR power-law slope & $\alpha$ & 1.0; 1.5; 2.0;  2.5\\
  \hline
  \end{tabular}
  \caption{{\bf List of the input parameters of the code CIGALE and their selected values
  (see Noll et al. 2009, for detail of parameters description).}}
  \label{table1}
\end{table*}

\section{The reliability of the SED fitting}
   {\bf For a SED fitting, the accuracy of the output relies on the input parameters. To give
proper physical parameters, such as SFR, the robustness of the results must be tested.
A straightforward method to check the reliability of the output of CIGALE
is to compare  with comprehensively known physical parameters.}

   In Fig. 1, we plot the difference in mJy between the
flux from the best  model estimated by CIGALE and the observed
flux for each broad band filter of W1, W2, W3 and W4. It shows that fluxes from the observation
are well reproduced by the code, though there is large dispersion. {\bf The typical error of model
flux increase with the wavelength from W1 to W4 induced by the increasing photometric sensitivity
 from W1 to W4(Cutri et al. 2012).

 There is a cloud of points in large flux region for W1 and W2 that
the model flux remain nearly unchanged  for the cloud of points  in large flux region for W1 and W2.
 This behaviour indicates  the less
fit quality for these three filters which is caused by relatively high uncertainties
in the object photometry. The relatively bright massive galaxies show greater discrepancies than dim ones, indicating the PSF
photometry is less reliable for brighter galaxies, because a considerable part of their flux is left
out by the relatively small beam size.}



\begin{figure*}[t]
\centering
\epsfxsize=6cm \epsfbox{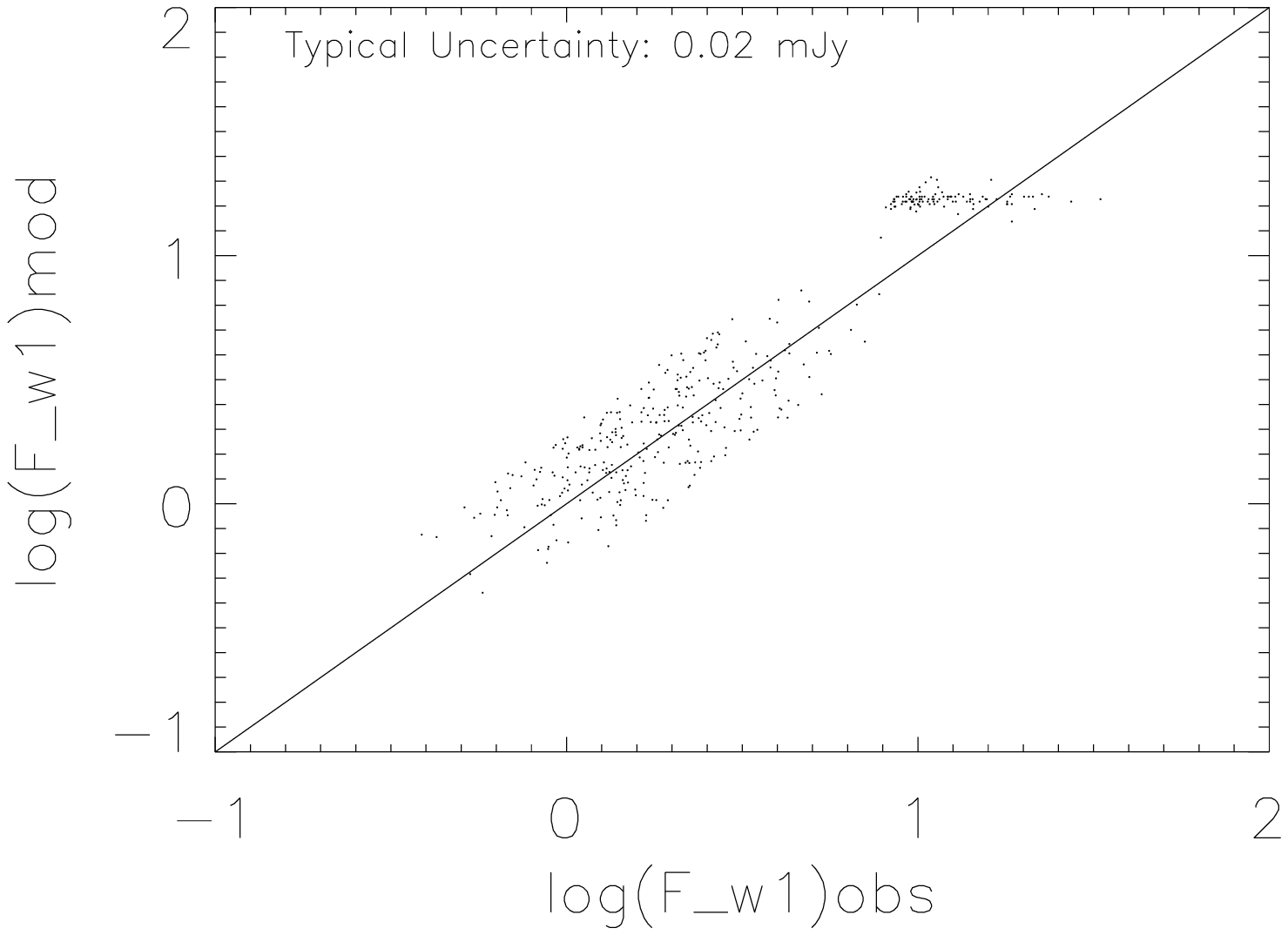}
\epsfxsize=6cm \epsfbox{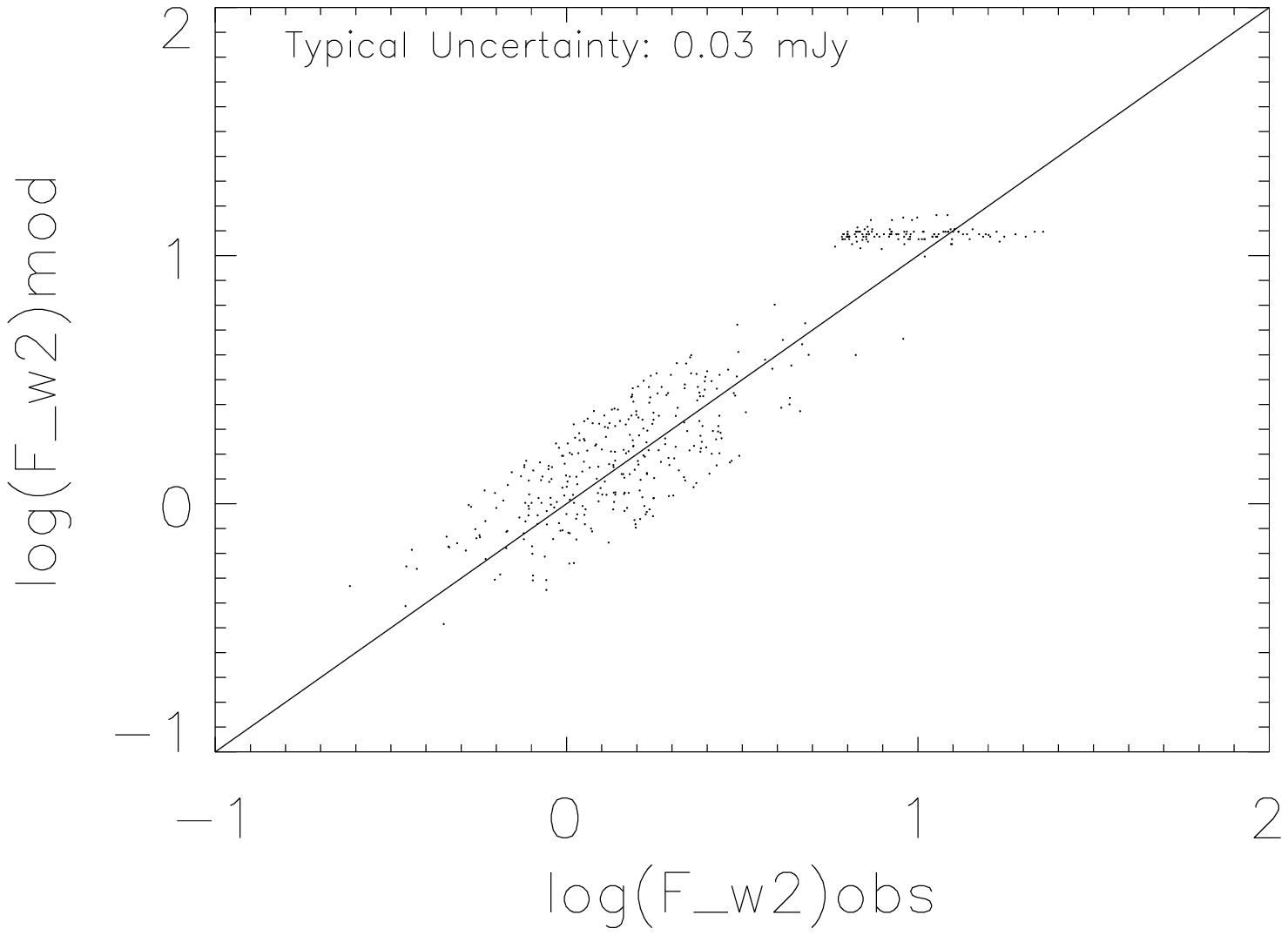}
\epsfxsize=6cm \epsfbox{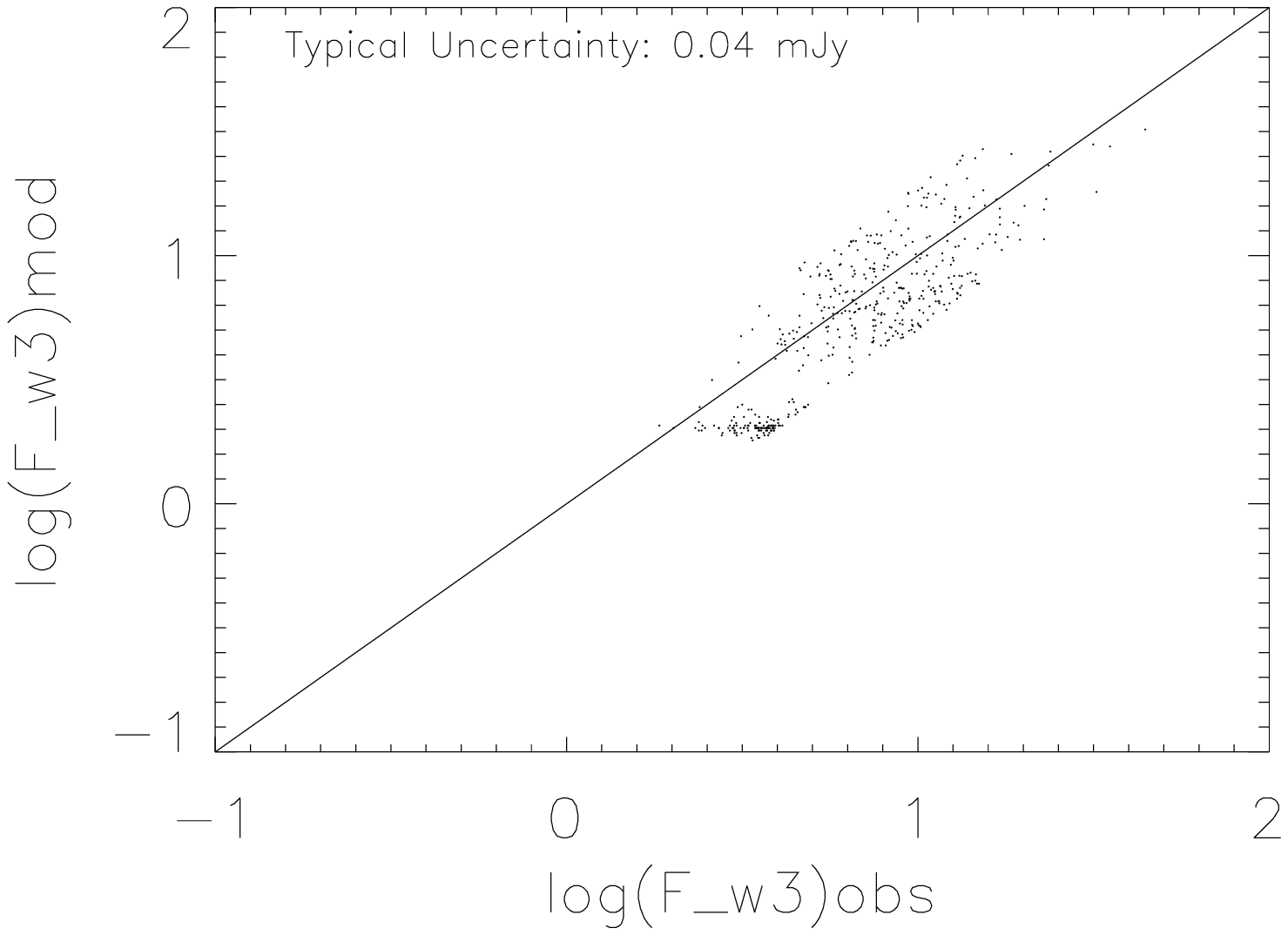}
\epsfxsize=6cm \epsfbox{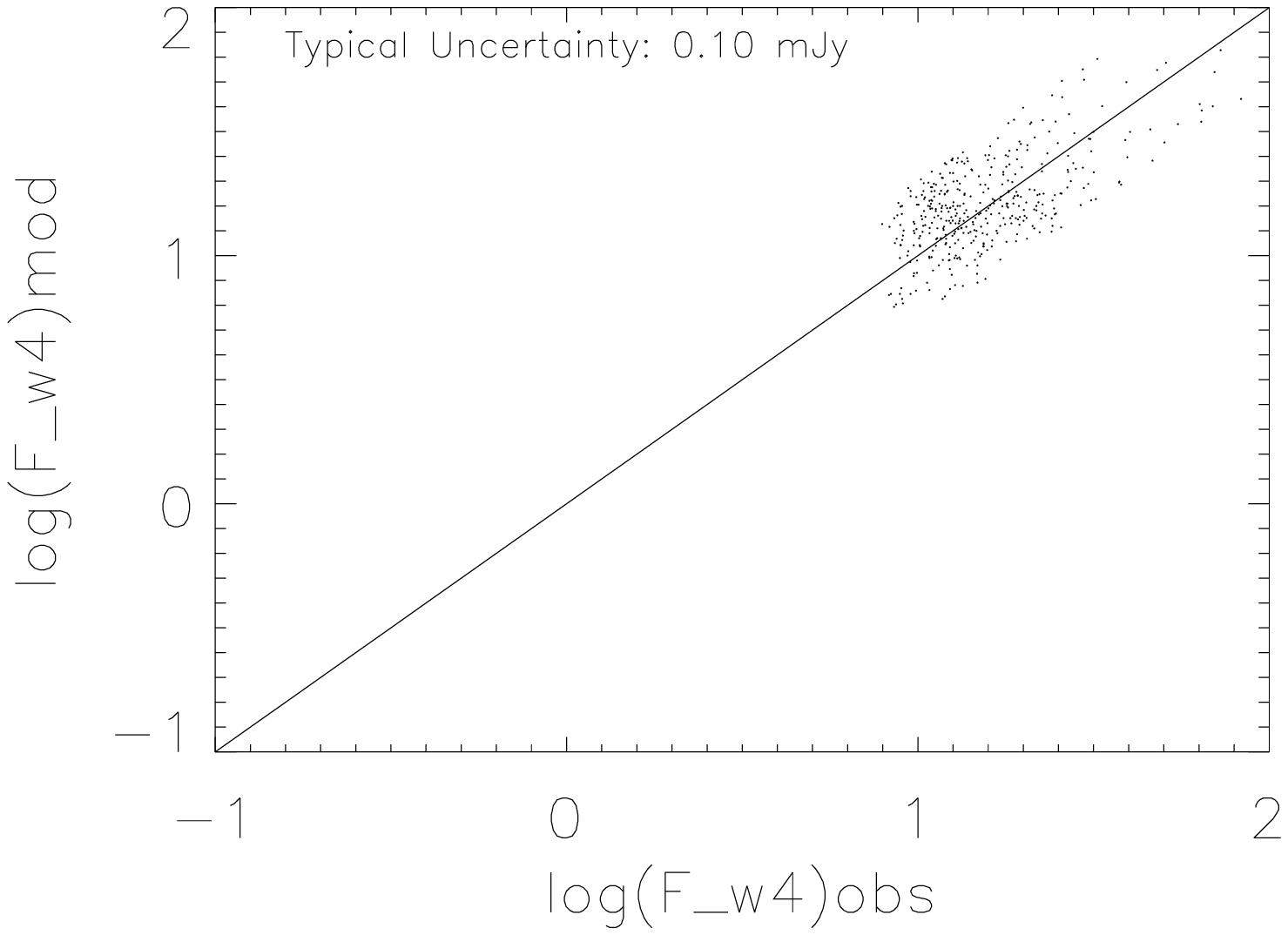}
\caption{Comparison  between the logarithmic flux from the best  model estimated by CIGALE and the observed
logarithmic flux from each broad band filter of W1, W2, W3 and W4.}
\label{fig-1}
\end{figure*}

  Though the code is able to well reproduce the observed data,
we cannot be content with  this result because it is possible that some
SEDs derived by CIGALE are degenerate and correspond to different parameter values.
We have to compare each parameter with its Bayesian estimation by the code and
the estimation by MPA. We only present the result of SFRs because
in this work the SFR is the only parameter in which we are interested.

In Fig. 2, we compare the SFR from  MPA($SFR_{mpa}$) and the instantaneous SFR from
CIGALE($SFR_m$) for our data sample.
$SFR_{mpa}$ are estimated using the galaxy photometry following Salim et al. (2007).
It shows that SFR values derived by MPA are 0.3 dex larger than
 the instantaneous SFR from CIGALE($SFR_m$) for our data sample below log($SFR_m$)=2, and there is
 a significant dispersion between them. The dispersion between $SFR_m$
and $SFR_{mpa}$  could be caused by an insufficient consideration of
thermally pulsating asymptotic giant branch (TP-AGB)
stars in older but widespread models of Bruzual \& Charlot
(2003) and Fioc \& Rocca-Volmerange (1997; PEGASE),
which were used to derive $SFR_{mpa}$ (Brinchmann et al. 2004). As a result,
it typically increases the stellar mass by 0.2 dex for star-forming galaxies
with  important star populations that have an intermediate age (Maraston et al. 2006).

{\bf This 0.3 dex offset also can be caused by the different choice of prior star formation history distributions
between  CIGALE and MPA. The choice of prior star formation history in  CIGALE
is a constant burst SFH for young stellar populations,
and an exponentially decreasing one for old stellar populations, while the prior distributions
used in MPA are designed to be as flat as possible.  Besides that,
the choice of star formation timescales in prior star formation history distributions
are also different  in this work(see Table 1) and in the MPA(Brinchmann et al. 2004).
Pacifici et al. 2012 emphasize that the physical parameters, such as
$SFR_{mpa}$,  highly depend on the choice of appropriate prior star formation history distributions
when applying the Bayesian approach to retrieve from observed spectral energy distributions of galaxies.

 The cloud of points above log($SFR_m$)=2 seems to lie closer to the MPA value.
 This agreement implies that the star formation timescales over which both SFR are calculated are similar for these
 most violent star-forming galaxies. Though the star formation history distribution used in MPA are designed to be flat
 on the whole, it becomes to decline steeply for the high SFR galaxies(Pacifici et al. 2012, their Fig.4c)
 which is similar to the  choice of  star formation history  of a constant burst SFH for young stellar populations plus
 an exponentially decreasing one for old stellar populations in  this work.
 }

\begin{figure*}
   \centering
   \includegraphics[scale=0.9,angle=0]{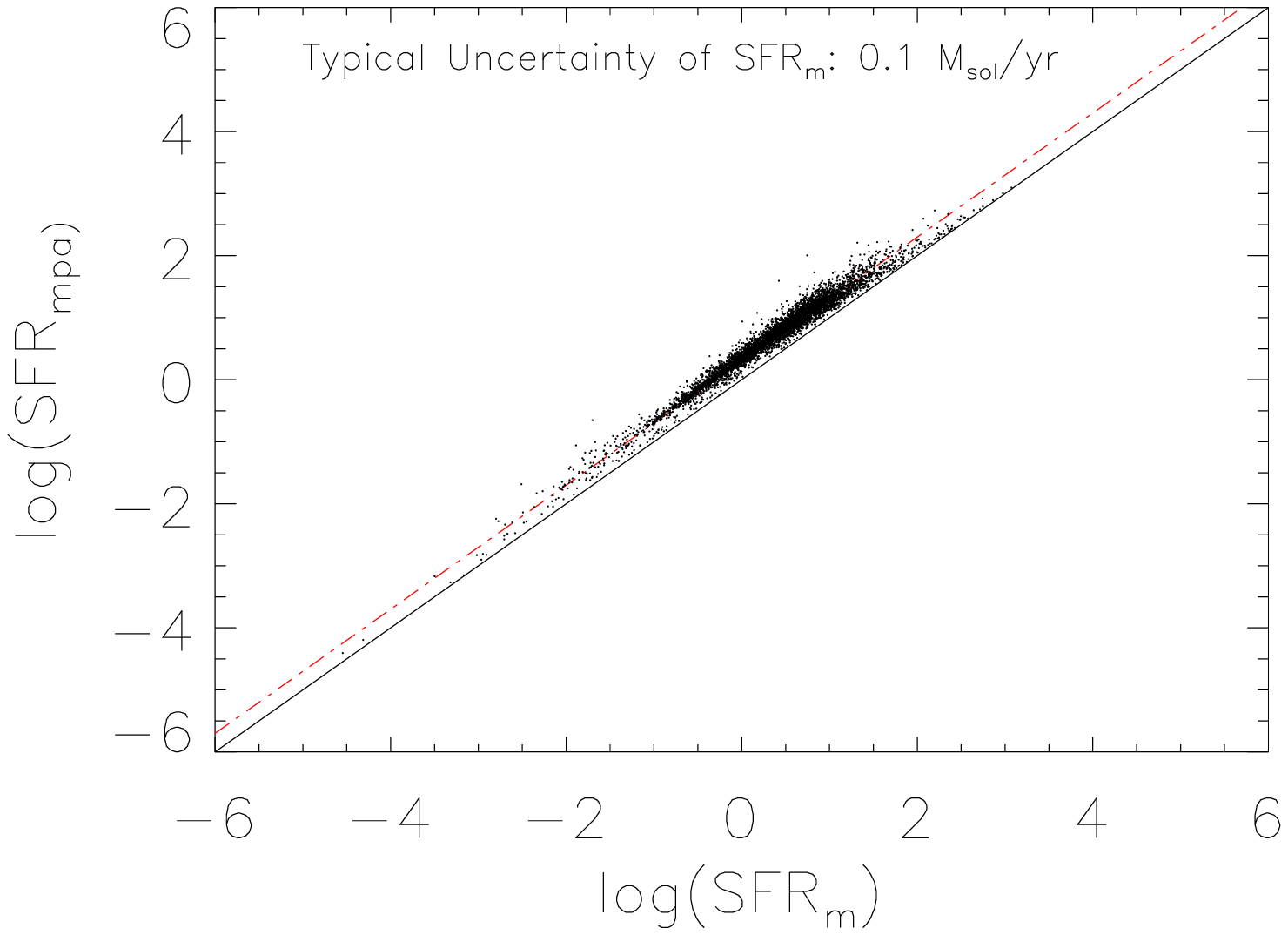}
   \caption{Comparison    between SFR from  MPA ($SFR_{mpa}$) and the instantaneous SFR from CIGALE ($SFR$) for our data sample.
   The logarithmic SFR is in units of M$_{\Zsolar}$/yr. }
              \label{Fig2}%
    \end{figure*}

 \section{ Star Formation Rates Calibrator}

Star-forming regions tend to be dusty and the
cross-section of dust absorption peaks in the UV. Radiation is then
reprocessed by dust and emerges beyond a few $\mu$m. As a result,
star formation should be closely related to $L_{\rm TIR}$.
{\bf This knowledge  is complicated by the complex physical conditions in the star-forming region.
For example, not all of the luminous energy produced by recently formed
stars are re-processed by dust in the infrared, which depends on the amount of dust.
Secondly, evolved stellar populations  also heat
the dust, which then emits in the FIR.
The other mechanisms are present in Calzetti et al. 2010 and references therein.}
As a result, it is necessary to check whether the
$L_{\rm TIR}$ can be a reliable SFR indicator.

To check whether $L_{\rm TIR}$ values are
reliable SFR indicators, In Fig. 3, we show the relation between $L_{\rm TIR}$
 and the instantaneous SFR from CIGALE for our data sample.
Although there is  {\bf average 0.5 dex  dispersion in the region of $-3.32 < log(SFR) < 3.07$},
all the objects merge into
a relatively tight, linear, and steep sequence, which gives
strong evidence that the IR luminosities
are good SFR indicators.

From the relation between SFR and $L_{\rm TIR}$  for our data sample in Figure 3,
we can define a  SFR calibration. The observed distribution of all the points in
this figure follows a linear least-squares fitting by the expression
shown as the dashed line in Fig. 3.

{\bf Since both SFR and  $L_{\rm TIR}$  are derived from the same fitting procedure,
 one may wonder whether the relationship is already present in the model library and what is the scatter.
 We plot the SFR- $L_{\rm TIR}$  relationship in Fig. 4 for all  galaxies, not only
 star-forming galaxies, but also AGN, low S/N LINER, and composite.
 We make this all  galaxies catalog by compiling a catalog of sources
 that have good detections from the WISE All-sky Survey,
 and then matching this  catalog  with the catalog of all galaxies in SDSS
   DR 9 provided by MPA(Section 2). There is $\sim$23,000 galaxies in total.

  It shows that  the relation between SFR and $L_{\rm TIR}$  for star-forming sample in Figure 3
  is nearly as same as it in Figure 4 for all galaxies. It imply that although the radiation mechanism of AGNs
  and LINER   is different from normal galaxies, AGN and LINER share a  SFR - $L_{\rm TIR}$ relation
  similar to normal star forming galaxy. A possible reason is that the
contribution to SED from AGN component and LINER is minor, therefore
the host galaxy component dominates the spectrum.

  The scatter in Figure 4 is $\sim$ 1.0dex, which is
  much larger than it in Figure 3. This larger scatter
  is obviously contributed by AGN activity and LINER.
  }

 In the next section, the dispersion in Fig.3 will be discussed.

\begin{figure*}
   \centering
   \includegraphics[scale=0.9,angle=0]{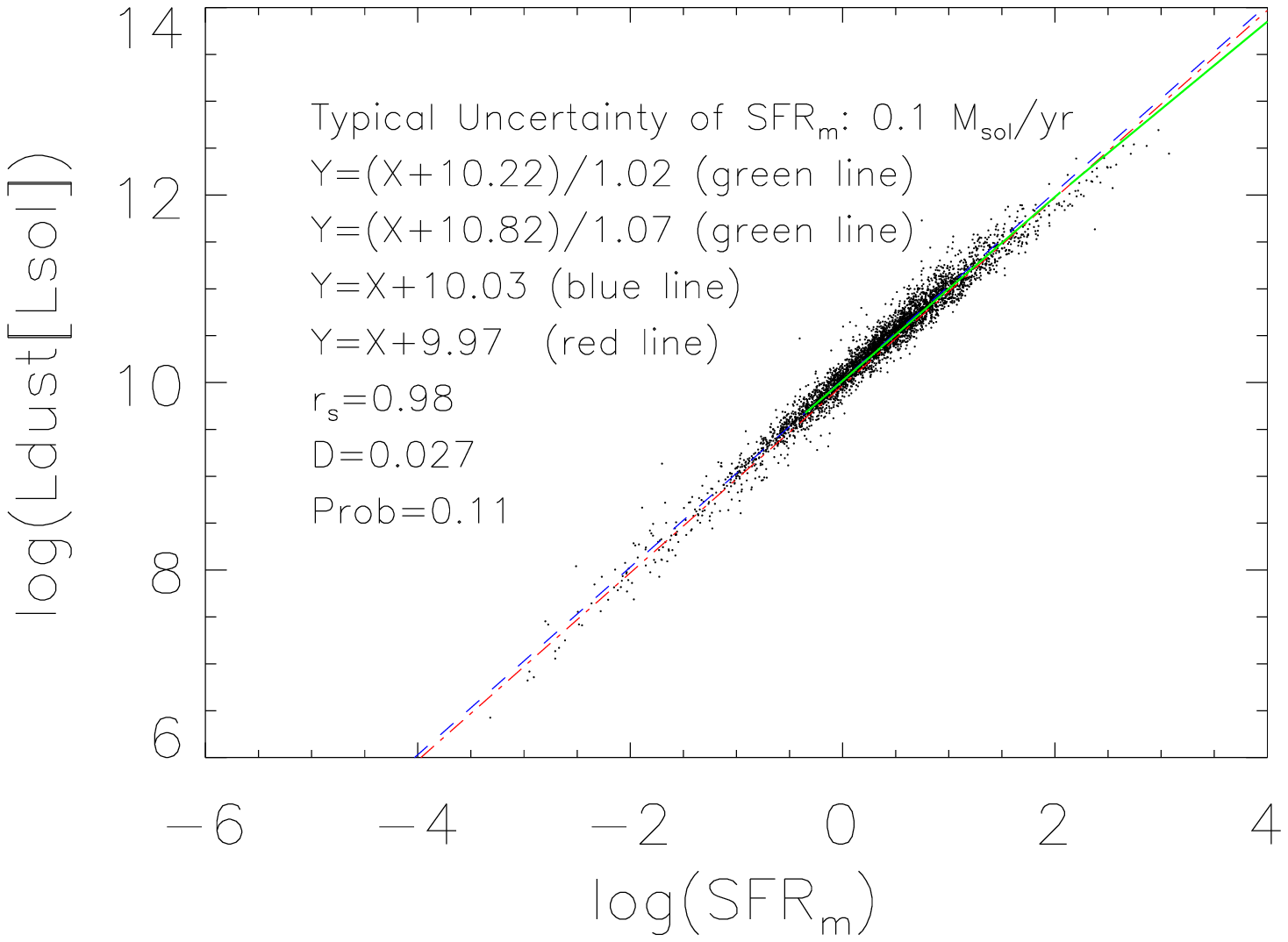}
   \caption{The relation between SFR and $L_{\rm TIR}$  for our data sample.
   The black dots use the instantaneous SFR from CIGALE($SFR_m$). The blue dashed line  denotes
the best-fit function.
The red dot-dashed lines are SFR calibrations from Buat et al. (2008).
The green lines are SFR calibrations from Rujopakarn et al.(2013).
The logarithmic SFR is in units of M$_{\Zsolar}$/yr. The logarithmic luminosity is in units of L$_{\Zsolar}$.
$r_s$ is the Spearman's correlation coefficient of the fit. D is maximum deviation between the cumulative
 distribution of the data and the fitted function.
 Prob is the significance level of   the two-sided Kolmogorov-Smirnov (K-S) statistic.}
 \label{Fig3}%
    \end{figure*}

\begin{figure*}
   \centering
   \includegraphics[scale=0.9,angle=0]{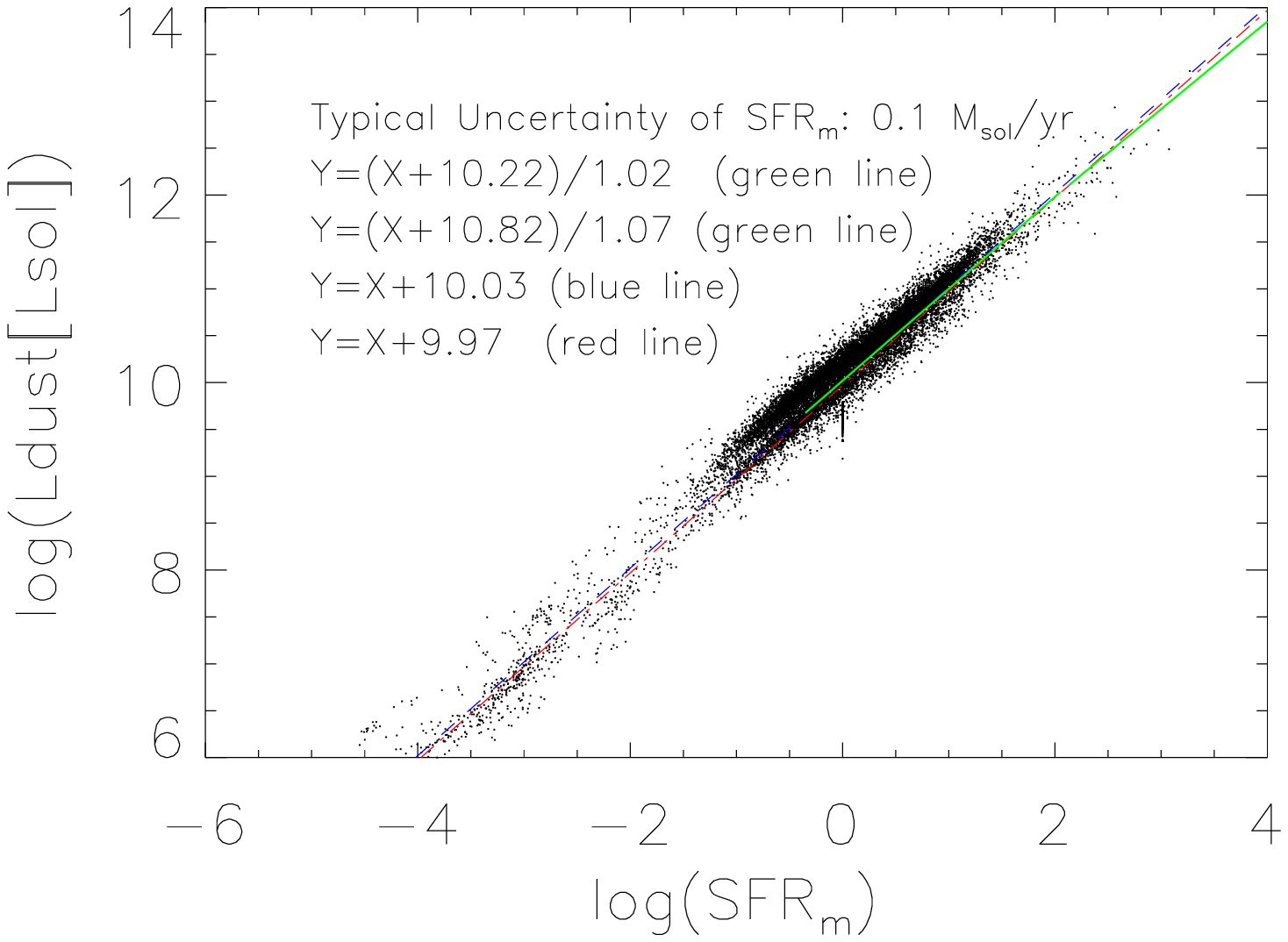}
   \caption{The relation between SFR and $L_{\rm TIR}$  for our data sample.
   The black dots use the instantaneous SFR from CIGALE($SFR_m$). The blue dashed line  denotes
the best-fit function in Fig.4.
The red dot-dashed lines are SFR calibrations from Buat et al. (2008).
The green lines are SFR calibrations from Rujopakarn et al.(2013).
The logarithmic SFR is in units of M$_{\Zsolar}$/yr. The logarithmic luminosity is in units of L$_{\Zsolar}$.
}
 \label{Fig4}%
    \end{figure*}

\section{ Discussion}
\subsection{Comparison with other SFR calibrations }

In Fig.3, we compare the SFR calibrations from Buat et al. (2008) and Rujopakarn et al.(2013) with ours.
The blue dashed line  denotes the best-fit function for our work.
The red dot-dashed lines are SFR calibrations from Buat et al. (2008).
{\bf The green lines are SFR calibrations from Rujopakarn et al.(2013)
(their Eq.7 and Eq.8, log(SFR) = 1.02log($L_{\rm TIR}$) - 10.22 where $-0.35< log(SFR) < 2.11$
and log(SFR) = 1.07log($L_{\rm TIR}$) - 10.82 where $log(SFR) >2.11$).
The three calibrations are consistent with each other on the whole besides slight offset. }

It shows that SFR values from  Buat et al. (2008) are $\sim$0.06 dex larger than ours.
The difference could be caused by Buat et al. (2008) assuming that  star formation history is a constant SFR over
$10^8$ years  and there is a Kroupa initial mass function ( Kroupa  2001) when making the calibration(their Eq 1),
while we use a Salpeter IMF and apply a constant burst SFH for young stellar populations
and an exponentially decreasing one for old stellar populations.
The use of a Salpeter IMF will cause the SFRs to be smaller by $\sim$0.18 dex than some other IMF with a more shallow
slope at low masses (Rieke et al. 2009).
Assuming a constant SFR over $10^8$ years  may not be proper because  the star formation
process has been proved to be far more complex than a single constant SFR (Kunth et al. 2000).

{\bf It shows that SFR calibrations from Rujopakarn et al.(2013) are
 more consistent with  Buat et al. (2008) than our calibration.
It can be understood that the calibration of Rujopakarn et al.(2013) is based on the same assumption as
Buat et al. (2008) that  a continuous star bursts and a Kroupa (2001) IMF.
}
\subsection{The origin of the Dispersion}

Although the SFRs are closely related to $L_{\rm TIR}$,
the dispersion of the relation is large. To further investigate the dispersion,
we plot the  relationship between the SFRs and  $L_{\rm TIR}$
in Fig.5 for different 4000{\AA} breaks, stellar masses, concentration values and metallicity.
{\bf To get coincident result,  we use the value of 4000{\AA} breaks, stellar masses, concentration values and metallicity
 from MPA. }
\begin{figure*}[t]
\centering
\epsfxsize=6cm \epsfbox{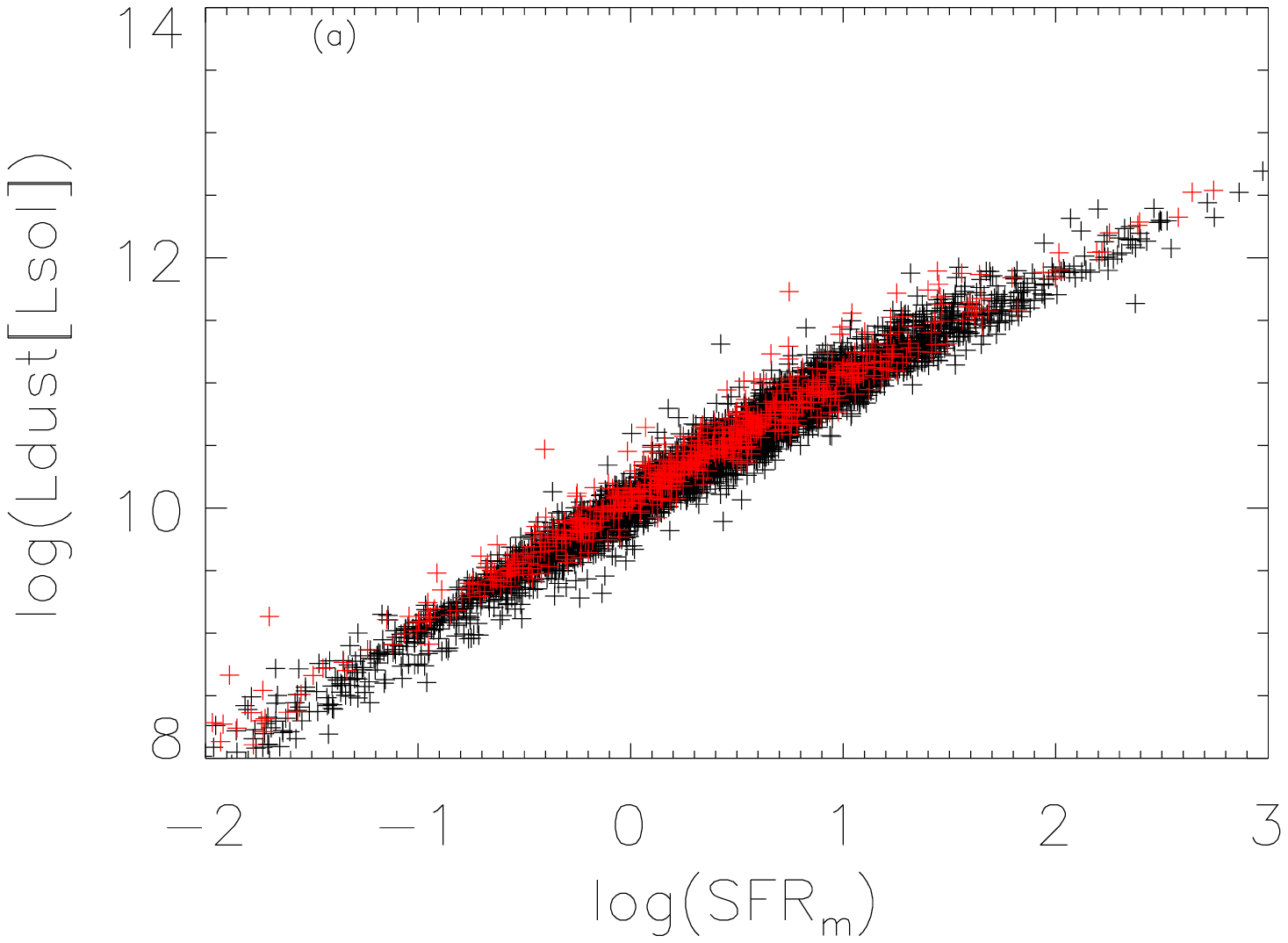}
\epsfxsize=6cm \epsfbox{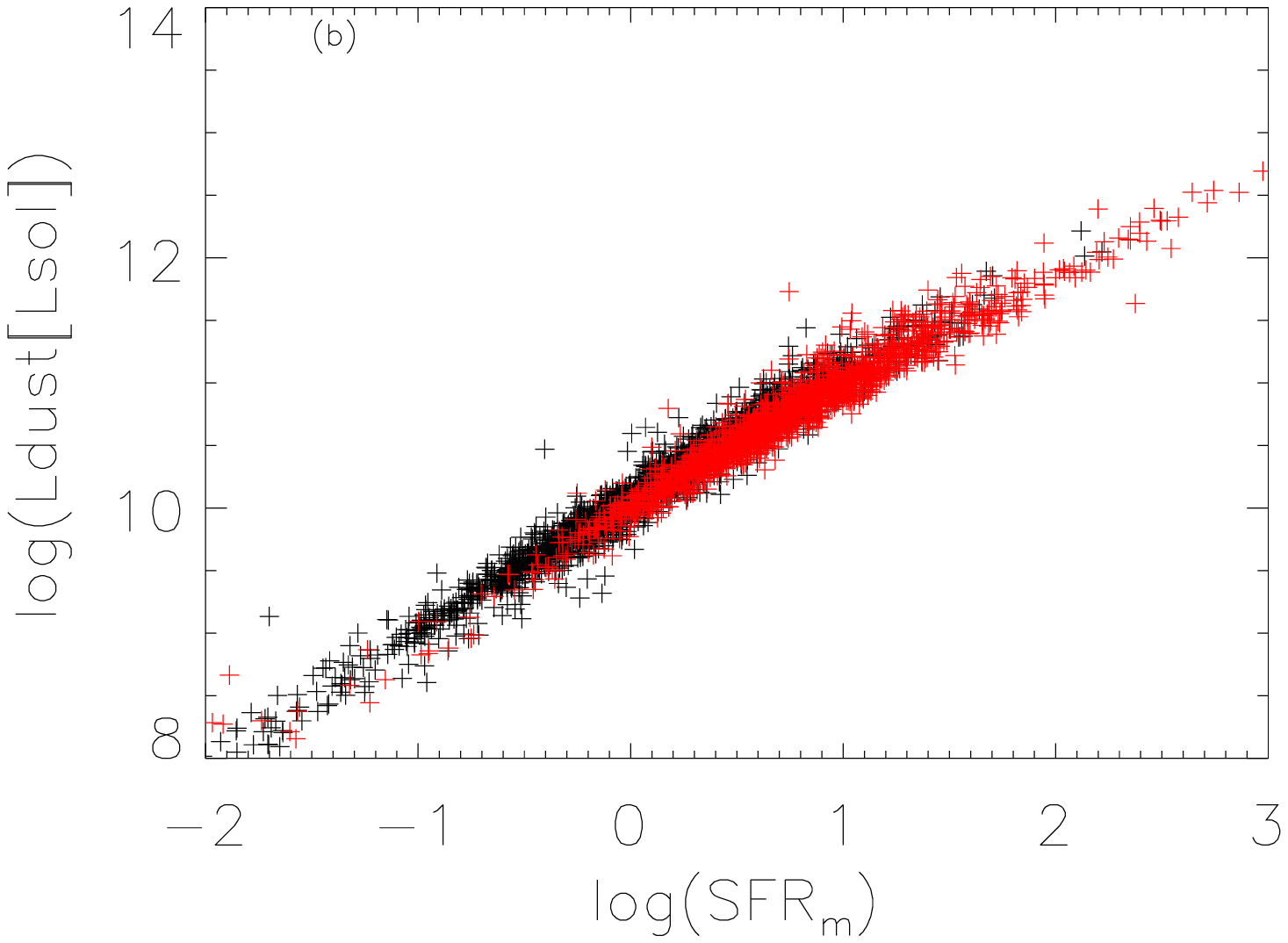}
\epsfxsize=6cm \epsfbox{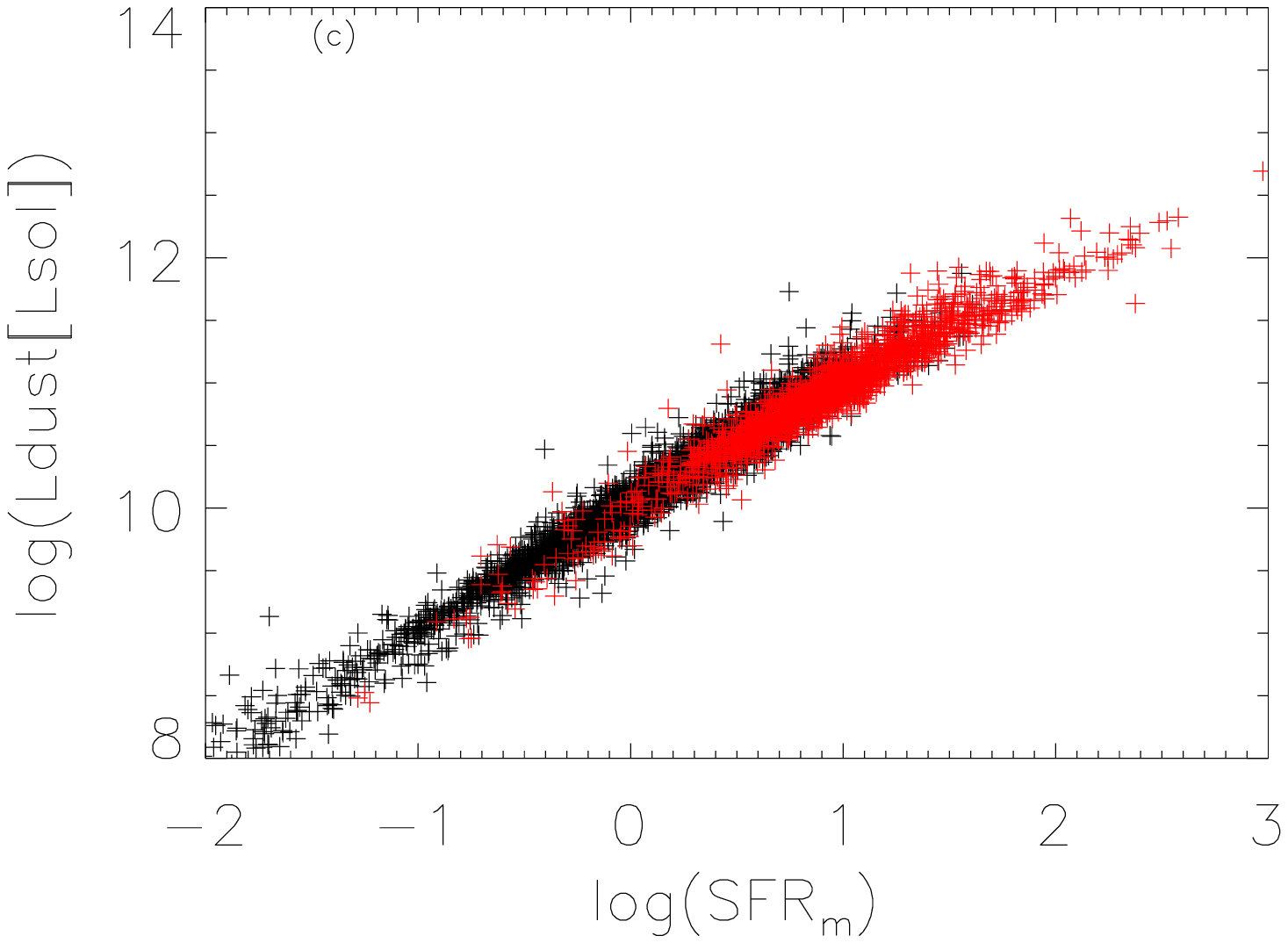}
\epsfxsize=6cm \epsfbox{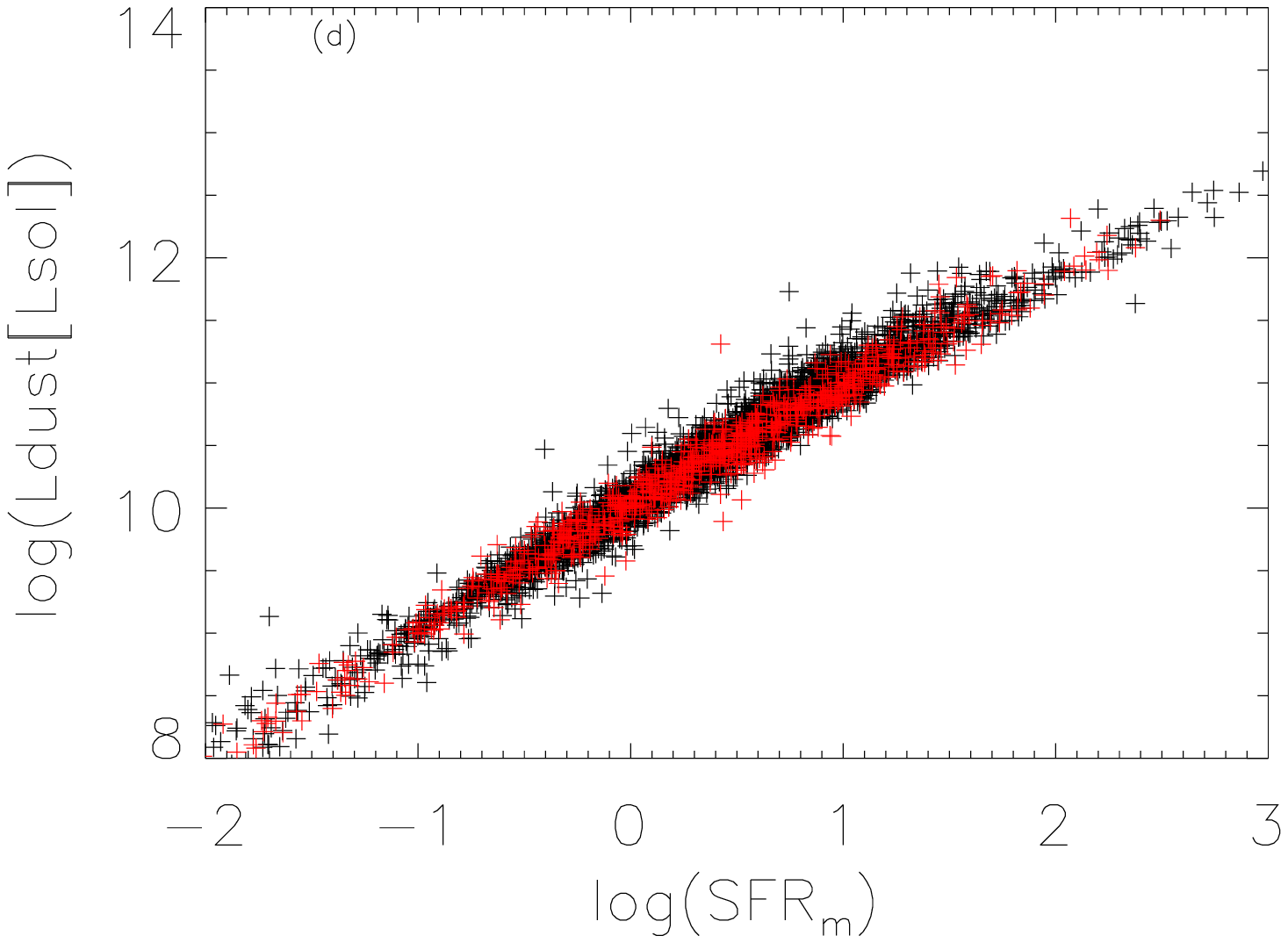}
\caption{The  relation between SFR and $L_{\rm TIR}$  for our data sample.
The sample in a) is subdivided into different 4000 {\AA} breaks.
Red and black crosses indicate the subsamples with the 4000{\AA} break $>$ 1.2 and
4000{\AA} break $<$ 1.2.
The sample in b) is subdivided into different metallicity.
Red and black dots indicate the subsamples of \hbox{$12\,+\,{\rm log(O/H)}$} $>$ 8.9 and
\hbox{$12\,+\,{\rm log(O/H)}$} $<$ 8.9.
The sample in c) is subdivided into different stellar masses.
Red and black crosses indicate the subsamples of log(stellar mass) $>$ 10.5 and
log(stellar mass) $<$ 10.5.
The sample in d) is subdivided into different concentration values.
Red and black crosses indicate the subsamples of $C > 2.6$ and
$C < 2.6$.
The logarithmic luminosity is in units of solar luminosity.
}
\label{fig-5}
\end{figure*}

\subsubsection{4000{\AA} Break }

To show the contribution of the star formation history to the
dispersion in the relation between SFR and $L_{\rm TIR}$,
we divided the data sample into  sub-samples
with  4000 {\AA} breaks less than 1.2 and larger than 1.2(Fig. 5a).
{\bf We choose 1.2 as cut-off point because the star formation density
 peaks at 4000 {\AA} breaks $\sim$ 1.2(Brinchmann \etal 2004). }
It shows that the $L_{\rm TIR}$-SFR relation at a low 4000 {\AA} break
is displaced to higher SFR and lower $L_{\rm TIR}$, especially in the low SFR region.
 We perform the two-sided Kolmogorov-Smirnov (K-S) statistic test for this case
and find that the significance level of  the test is 0.14, supporting
the idea that the dispersion is partly related to the 4000 {\AA} break.

Brinchmann \etal(2004) show that most of the star formation takes
place in galaxies with a low 4000 {\AA} break.
This can explain  the behavior of the low 4000 {\AA} break displaced
to higher SFR, where galaxies with a low 4000 {\AA} break have a
younger stellar population, and therefore higher SFR values than
high 4000 {\AA} break galaxies with the same SFR.
The selection effects also contribute to the effect that galaxies with a high 4000
{\AA} break tend to be excluded from our data sample
because these galaxies usually have old stellar populations and
seldom show emission lines. If one high 4000 {\AA} break
galaxy indeed shows emission lines, it will tend to have
less starburst activity, and hence a lower SFR.

\subsubsection{ Metallicity, Stellar mass and Concentration Index }

It is interesting to study whether metallicity of galaxies are related to SFR or not. To
investigate this more clearly, we divided the data sample into  sub-samples
with different metallicity (Fig. 5b).
The SFR and $L_{\rm TIR}$ relation at low oxygen abundance is displaced to lower SFR and lower luminosity.
We can explain this because the majority of low oxygen abundance galaxies have less star forming
events, therefore they have relatively lower SFR.
The behavior that the higher oxygen abundance galaxies tends to have higher luminosity (mass) has been
studied by many authors (Shi et al. 2005, and references therein ). With the luminosity and metallicity
correlation, the most straightforward interpretation
 is that more massive galaxies form fractionally more stars in a Hubble time (higher luminosity)
 than low-mass galaxies, and then have higher metallicity.
 We perform the two-sided Kolmogorov-Smirnov (K-S) statistical test for this case
and find that the significance level of  the test is 0.

 The stellar mass distribution of SDSS galaxies peaks at $log M* \sim 10.5$, so we
 divided the data sample into  sub-samples with $log M* < or >10.5$ in Fig. 5c.
 It is obvious that the $L_{\rm TIR}$-SFR relation at low stellar mass ($log M* < 10.5$ )is displaced to
lower SFR and lower luminosity. We can explain it that   there
is a strong positive correlation between stellar mass and metallicity (Brinchmann \etal 2004),
so stellar mass should have the same behaviour as metallicity for the $L_{\rm TIR}$-SFR relation.
We perform the two-sided Kolmogorov-Smirnov (K-S) statistical test for this case
and find that the significance level of  the test is 0.

The galaxies with the concentration value $C > 2.6$ are mostly early type galaxies
whereas late type galaxies have $C < 2.6$ . It is well known
that early type galaxies are dominated by old/small mass stars, but late type galaxies
are dominated by young/massive stars(Kauffmann et al. 2003).
 In Fig. 5d, we subdivide the sample into sub-samples with different C.
It is clear that the concentration value does not contribute
to the dispersion. We perform the two-sided Kolmogorov-Smirnov (K-S) statistical test for this case
and find that the significance level of  the test is 3.2E-5.
\section{Conclusions}

{\bf We have collected  a large sample of star-forming galaxies, by matching the WISE data with
 star forming galaxies catalog of the SDSS DR9.}
We estimated the global properties of this sample such as the star formation rate, the total infrared
luminosity, the stellar mass, the 4000{\AA} break with the code CIGALE
which performs a Bayesian analysis to deduce how these parameters are related
to the dust attenuation and star formation of each galaxy.
Regression analysis was conducted to
investigate $L_{\rm TIR}$-SFR relations for this sample.
We have confirmed the existence of the correlation between the $L_{\rm TIR}$
and SFR  and have obtained a calibration that can
be used as a method for determining SFR for star-forming galaxies. {\bf The calibration
holds for star forming galaxies in the redshift range of less than 0.1, $-3.32 < log(SFR) < 3.07$,
$0.35 <log( L_{\rm TIR}) <12.7$.}
The standard Kolmogorov-Smirnov test shows that
the dispersion and non-linearity in the
relation between $L_{\rm TIR}$  and SFR is partly related
to the galaxy's properties,  such as the 4000 {\AA} break.

\acknowledgments{
 This work was funded by the National Natural Science Foundation of
China (NSFC) (Grant Nos.~11203001 and 10873012), the National Basic Research Program
 of China (973 Program) (Grant No.~2007CB815404), and the Chinese
Universities Scientific Fund (CUSF).

This publication makes use of data products from the Wide-field Infrared
Survey Explorer, which is a joint project of the University of California,
Los Angeles, and the Jet Propulsion Laboratory/California Institute of
Technology, funded by the National Aeronautics and Space Administration.
Funding for the Sloan Digital Sky Survey (SDSS) has been provided
by the Alfred P. Sloan Foundation, the Participating Institutions, the
National Aeronautics and Space Administration, the National Science
Foundation, the U.S. Department of Energy, the Japanese Monbukagakusho,
and the Max Planck Society.}

\appendix



\begin{thebibliography}{}

\bibitem[Alonso--Herrero et al.(2006)]{AlonsoHerrero2006} Alonso--Herrero, A., Rieke, G.H., Rieke, M.J., Colina, L., Perez-Gonzalez, P.G., \& Ryder, S.D. 2006, \apj, 650, 835

\bibitem[Balogh et al.(1998)]{1998ApJ...504L..75B} Balogh, M.~L., Schade,
D., Morris, S.~L., Yee, H.~K.~C., Carlberg, R.~G.,
\& Ellingson, E.\ 1998, \apjl, 504, L75

\bibitem[Buat et
al.(2008)]{2008A&A...483..107B} Buat, V., Boissier, S., Burgarella, D., et al.\ 2008, \aap, 483, 107

\bibitem[Buat et
al.(2011)]{2011A&A...529A..22B} Buat, V., Giovannoli, E., Takeuchi, T.~T., et al.\ 2011, \aap, 529, A22

\bibitem[Bigiel et al.(2008)]{Bigiel2008} Bigiel, F., Leroy, A., Walter, F., Brinks, E., de Blok, W.J.G.,  Madore, B., Thornley, M.D. 2008, \aj, 136, 2846

\bibitem[Brinchmann et al.(2004)]{2004MNRAS.351.1151B} Brinchmann, J.,
Charlot, S., White, S.~D.~M., Tremonti, C., Kauffmann, G., Heckman, T.,
\& Brinkmann, J.\ 2004, \mnras, 351, 1151

\bibitem[Bruzual
\& Charlot(2003)]{2003MNRAS.344.1000B} Bruzual, G., \& Charlot, S.\ 2003, \mnras, 344, 1000

\bibitem[Buat et
al.(2002)]{2002A&A...383..801B} Buat, V., Boselli, A., Gavazzi, G., \& Bonfanti, C.\ 2002, \aap, 383, 801


\bibitem[Buat et al.(2005)]{2005ApJ...619L..51B} Buat, V.,
Iglesias-P{\'a}ramo, J., Seibert, M., et al.\ 2005, \apjl, 619, L51


\bibitem[Calzetti et al.(1994)]{1994ApJ...429..582C} Calzetti, D., Kinney,
A.~L., \& Storchi-Bergmann, T.\ 1994, \apj, 429, 582


\bibitem[Calzetti et al.(2000)]{calze00} Calzetti, D., Armus,
L., Bohlin, R.~C., Kinney, A.~L., Koornneef, J.,
\& Storchi-Bergmann, T.\ 2000, \apj, 533, 682

\bibitem[Calzetti et al.(2005)]{Calzetti2005} Calzetti, D., Kennicutt, R.C.,
Bianchi, L., Thilker, D.A., Dale, D.A., Engelbracht, C.W., Leitherer, C.,
Meyer, M.J., et al. 2005, \apj, 633, 871

\bibitem[Calzetti et al.(2007)]{2007ApJ...666..870C} Calzetti, D.,
Kennicutt, R.~C., Engelbracht, C.~W., et al.\ 2007, \apj, 666, 870

\bibitem[Calzetti et al.(2009)]{2009PASP..121..937C} Calzetti, D., \& Kennicutt, R.~C.\ 2009, \pasp, 121, 937



\bibitem[Calzetti et al.(2010)]{2010ApJ...714.1256C} Calzetti, D., et al.\
2010, \apj, 714, 1256


 \bibitem[Charlot, S, Fall, S. Michael\/(2000)]{CF2000}
Charlot, S, Fall, S. M.\ 2000, ApJ, 539, 718C

\bibitem[Chary
\& Elbaz(2001)]{2001ApJ...556..562C} Chary, R., \& Elbaz, D.\ 2001, \apj, 556, 562

\bibitem[Cutri et al.(2012)]{2012wise.rept....1C} Cutri, R.~M., Wright,
E.~L., Conrow, T., et al.\ 2012, Explanatory Supplement to the WISE All-Sky
Data Release Products, 1

\bibitem[da Cunha et al.(2008)]{2008MNRAS.388.1595D} da Cunha, E., Charlot,
S., \& Elbaz, D.\ 2008, \mnras, 388, 1595


\bibitem[Dale
\& Helou(2002)]{2002ApJ...576..159D} Dale, D.~A., \& Helou, G.\ 2002, \apj, 576, 159


\bibitem[Fioc
\& Rocca-Volmerange(1997)]{1997A&A...326..950F} Fioc, M., \& Rocca-Volmerange, B.\ 1997, \aap, 326, 950

\bibitem[Kauffmann et al.(2003)]{2003MNRAS.346.1055K}
Kauffmann, G., et al.\ 2003, \mnras, 346, 1055

\bibitem[Kennicutt(1998)]{1998ARA&A..36..189K}
Kennicutt, R.~C.\ 1998, \araa, 36, 189

\bibitem[Kennicutt et al.(2007)]{Kennicutt2007} Kennicutt, R.C., Calzetti, D., Walter, F., Helou, G.,m Hollenbach, D., Armus, L., Bendo, G., Dale, D.A., Draine, B.T., Engelbracht, C.W., et al. 2007a, \apj, 671, 333

\bibitem[Kong(2004)]{2004A&A...425..417K} Kong, X.\ 2004, \aap, 425, 417


\bibitem[Kroupa(2001)]{2001MNRAS.322..231K} Kroupa, P.\ 2001, \mnras, 322,
231

\bibitem[Kunth {\&Ouml}stlin(2000)]{2000A&ARv..10....1K} Kunth, D., {\&Ouml}stlin, G.\ 2000, \aapr, 10, 1


\bibitem[Lagache et al.(2003)]{2003MNRAS.338..555L} Lagache, G., Dole, H.,
\& Puget, J.-L.\ 2003, \mnras, 338, 555

\bibitem[Mann et al.(2002)]{2002MNRAS.332..549M} Mann, R.~G., et al.\ 2002,
\mnras, 332, 549

\bibitem[Maraston(2005)]{2005MNRAS.362..799M} Maraston, C.\ 2005, \mnras,
362, 799

\bibitem[Maraston et al.(2006)]{2006ApJ...652...85M} Maraston, C., Daddi,
E., Renzini, A., et al.\ 2006, \apj, 652, 85

\bibitem[Marcillac et
al.(2006)]{2006A&A...451...57M} Marcillac, D., Elbaz, D., Chary, R.~R., Dickinson, M., Galliano, F., \& Morrison, G.\ 2006, \aap, 451, 57

\bibitem[Noll et
al.(2009)]{2009A&A...507.1793N} Noll, S., Burgarella, D., Giovannoli, E., Buat, V., Marcillac, D., \& Mu{\~n}oz-Mateos, J.~C.\ 2009, \aap, 507, 1793

\bibitem[Nordon et
al.(2010)]{2010A&A...518L..24N} Nordon, R., Lutz, D., Shao, L., et al.\ 2010, \aap, 518, L24

\bibitem[Pacifici et al.(2012)]{2012MNRAS.421.2002P} Pacifici, C., Charlot,
S., Blaizot, J., \& Brinchmann, J.\ 2012, \mnras, 421, 2002

\bibitem[Persic \& Rephaeli(2007)]{Persic2007} Persic, M., \& Rephaeli, Y. 2007, \aap, 463, 481

\bibitem[Pilyugin et
al.(2004)]{2004A&A...423..427P} Pilyugin, L.~S., Contini, T., \& V{\'{\i}}lchez, J.~M.\ 2004, \aap, 423, 427


\bibitem[Rieke et al.(2009)]{2009ApJ...692..556R} Rieke, G.~H.,
Alonso-Herrero, A., Weiner, B.~J., P{\'e}rez-Gonz{\'a}lez, P.~G., Blaylock,
M., Donley, J.~L., \& Marcillac, D.\ 2009, \apj, 692, 556


\bibitem[Rosa--Gonzalez et al.(2007)]{Rosa2007} Rosa--Gonzalez, D., Burgarella, D., Nandra, K., Kunth, D., Terlevich, E., \& Terlevich, R. 2007, \mnras, 379, 357

\bibitem[Rujopakarn et al.(2013)]{2013ApJ...767...73R} Rujopakarn, W.,
Rieke, G.~H., Weiner, B.~J., et al.\ 2013, \apj, 767, 73

\bibitem[Salim et al.(2007)]{2007ApJS..173..267S} Salim, S., et al.\ 2007,
\apjs, 173, 267

\bibitem[Schmitt et al.(2006)]{Schmitt2006} Schmitt, H.R., Calzetti, D., Armus, L., Giavalisco, M., Heckman, T.M., Kennicutt, R.C., Leitherer, C.,  \& Meurer, G.R. 2006, \apjs, 164, 52

\bibitem[Serra et al.(2011)]{2011ApJ...740...22S} Serra, P., Amblard, A.,
Temi, P., et al.\ 2011, \apj, 740, 22

\bibitem[Shi et
al.(2005)]{2005A&A...437..849S} Shi, F., Kong, X., Li, C., \& Cheng, F.~Z.\ 2005, \aap, 437, 849


\bibitem[Stanghellini et al.(2007)]{2007ApJ...671.1669S} Stanghellini, L.,
Garc{\'{\i}}a-Lario, P., Garc{\'{\i}}a-Hern{\'a}ndez, D.~A.,
Perea-Calder{\'o}n, J.~V., Davies, J.~E., Manchado, A., Villaver, E.,
\& Shaw, R.~A.\ 2007, \apj, 671, 1669


\bibitem[Wright et al.(2010)]{2010AJ....140.1868W} Wright, E.~L., et al.\
2010, \aj, 140, 1868

\bibitem[York et al.(2000)]{2000AJ....120.1579Y} York, D.~G., Adelman, J.,
Anderson, J.~E., Jr., et al.\ 2000, \aj, 120, 1579

\bibitem[Zheng et al.(2007)]{2007ApJ...670..301Z} Zheng, X.~Z., Dole, H.,
Bell, E.~F., Le Floc'h, E., Rieke, G.~H., Rix, H.-W.,
\& Schiminovich, D.\ 2007, \apj, 670, 301

\bibitem[Zhu et al.(2008)]{Zhu2008} Zhu, Y.-N., Wu, H., Cao, C., \& Li, H.-N. 2008, \apj, 686, 155
\end{thebibliography}
\end{document}